# A Comprehensive Study on Medical Image Segmentation using Deep Neural Networks

Loan Dao, Ngoc Quoc Ly

Dept. of Computer Vision and Cognitive Cybernetics
University of Science, VNUHCM, Ho Chi Minh, Vietnam
Viet Nam National University, Ho Chi Minh City, Vietnam

*Abstract*—Over the past decade, Medical Image Segmentation (MIS) using Deep Neural Networks (DNNs) has achieved significant performance improvements and holds great promise for future developments. This paper presents a comprehensive study on MIS based on DNNs. Intelligent Vision Systems are often evaluated based on their output levels, such as Data, Information, Knowledge, Intelligence, and Wisdom (DIKIW), and the state-of-the-art solutions in MIS at these levels are the focus of research. Additionally, Explainable Artificial Intelligence (XAI) has become an important research direction, as it aims to uncover the "black box" nature of previous DNN architectures to meet the requirements of transparency and ethics. The study emphasizes the importance of MIS in disease diagnosis and early detection, particularly for increasing the survival rate of cancer patients through timely diagnosis. XAI and early prediction are considered two important steps in the journey from "intelligence" to "wisdom." Additionally, the paper addresses existing challenges and proposes potential solutions to enhance the efficiency of implementing DNN-based MIS.

*Keywords*—*Medical image segmentation (MIS); SOTA solutions in MIS; XAI; early disease diagnosis*

## I. INTRODUCTION

Computers store images as grids of pixels, each containing a color value. These digital images are considered unstructured data, and image segmentation is the process of partitioning pixels into separate regions that correspond to a single object or class. This is accomplished by labeling each pixel with its corresponding class.

Image segmentation provides a deeper understanding of the structure of an image and is a crucial processing step in many image and video applications. It serves as the foundation for other challenges, such as object detection, image classification, and image analysis.

In medical imaging, image segmentation involves separating organs, disease regions, tumors, or anomalies to assist in diagnosis, detect pathology, and monitor the progression of diseases. MIS is a challenging task because of the slow grayscale variation in medical images, making it difficult to distinguish objects.

Currently, AI applications have reached the "Intelligence" level in DIKIW [1], which stands for Data, Information, Knowledge, Intelligence, and Wisdom. In MIS, "Intelligence" is demonstrated through highly accurate segmentation results, even on low-contrast, blurry, noisy images [2]. Additionally,

"Intelligence" in MIS is not only demonstrated through the ability to segment organs, but also through the ability to segment lesions [2].

"Wisdom", the highest level in the DIKIW hierarchy, represents humanity's ultimate goal. In MIS systems, the output not only provides segmented medical images but also offers an explanation for the segmentation results through eXplainable Artificial Intelligence (XAI). This helps to increase trust in the results among both doctors and patients, satisfying the demands for transparency and medical ethics, which is a critical aspect in realizing the practical application of AI-based disease diagnosis. Transparent segmentation results also contribute to early disease diagnosis, leading to improved treatment, monitoring, and healthcare processes. In healthcare systems, predicting future health conditions to plan appropriate care and potentially reducing the risk of death holds significant humanitarian significance. XAI and early prediction are two crucial steps towards bridging the "Intelligence-to-Wisdom" gap.

The following major contributions are presented:

- The state-of-the-art solutions in MIS focus on key factors such as network architecture, data, loss function, and evaluation metrics. The paper specifically explains the development process of the network architecture from the perspective of three levels of an intelligent vision system: the backbone, the typical network architecture, and applications of MIS.

- The paper focuses on the state-of-the-art solutions in MIS and highlights the current interest in XAI-based MIS to meet ethical and legal requirements.

- The paper presents a new perspective on MIS by incorporating the capability of making early predictions based on the results, which contributes to the improvement of community healthcare systems.

The rest of the paper is organized as follows: In Section II the state-of-the-art solutions for deep neural network-based medical image segmentation are presented. Section III focuses on the explanation of black-box models using eXplainable Artificial Intelligence (XAI) to increase trust among end-users. The paper analyzes early prediction techniques for disease progression in Section IV. The challenges and proposed solutions for improving the efficiency of future clinical applications are discussed in Section V. Finally, the conclusions of this comprehensive study and suggestions for future research are presented in Section VI.

## II. STATE-OF- THE-ART SOLUTIONS IN MEDICAL IMAGE SEGMENTATION

This section presents state-of-the-art solutions in MIS, including network architectures, data, loss functions, and evaluation metrics. The specifics are depicted in Fig. 1.

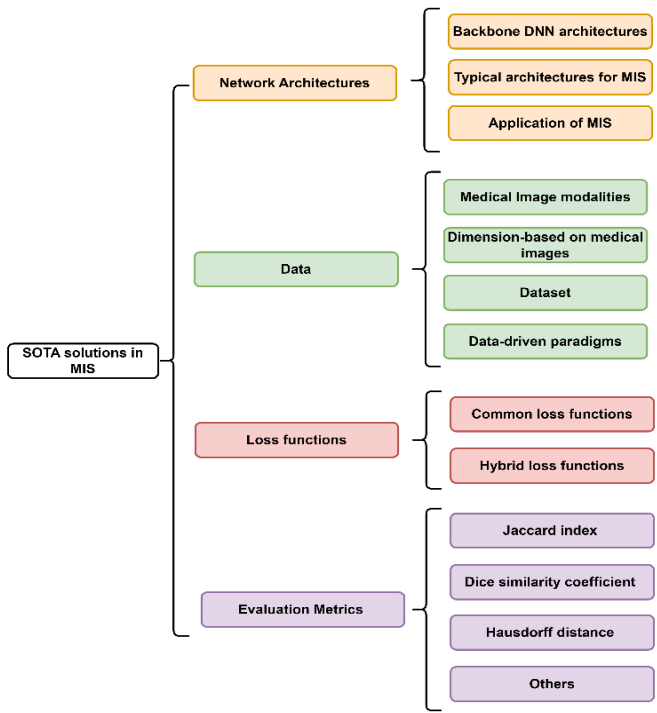

Fig. 1. The pipeline of SOTA solutions in MIS.

### A. Network Architectures

This study conducts a survey of state-of-the-art solutions in MIS based on the standard framework of intelligent vision systems (IVS). An IVS encompasses three main levels.

Level 1 encompasses the backbone deep neural network (DNN) architectures used in image segmentation.

Level 2 builds on Level 1 to create specialized image segmentation models.

Level 3 leverages the knowledge from Levels 1 and 2 to develop typical applications for medical image segmentation.

This study thoroughly describes and explains the three levels in the IVS framework (TABLE I. )

In the following sections, typical modules and architectures for each level will be thoroughly examined.

Level 1: Background DNN architectures:

This section introduces typical background neural network architectures and state-of-the-art modules aimed at enhancing the efficiency of segmentation.

*1) The typical background network architectures:* The paper delves into several typical background architectures, including Convolutional Neural Networks (CNNs), Recurrent Neural Networks (RNNs), Graph Neural Networks (GNNs), Generative Adversarial Networks (GANs), and Transformers.

- Convolutional neural network (CNN) (1982, [3]). Before gaining popularity, Convolutional Neural Networks (CNNs) have gone through several historical stages. The first foundation for convolution was laid by K. Fukushima et al. through a series of works, including "Cognitron: A self-organizing multilayered neural network" (1975), "Neural network model for a mechanism of pattern recognition unaffected by shift in position-Neocognitron" (1979), "Neocognitron: A new algorithm for pattern recognition tolerant of deformations and shifts in position" (1982). Since neighboring pixels in an image typically have strong inter-dependencies, K. Fukushima introduced the concept of "connectable areas" to extract features in "neighborhoods" instead of fully connecting the layers. This marked the first paradigm for unsupervised pattern recognition. The breakthrough of Convolutional Neural Networks (CNNs) lies in the use of two-dimensional filters, which are capable of extracting meaningful features from locally connected subsamples. These filters have smaller size, resulting in more optimal computing and storage capacity compared to earlier fully connected networks. The output from the filters is fed into an activation function, which adds non-linearity to the feature space and can be learned during training. Additionally, the non-linear activation function generates outputs that are typically monitored through subsampling. This allows for the aggregation of outputs, making the input insensitive to geometric deviations such as distortion, resizing, and repositioning of the sample input. CNNs have been successful in overcoming challenges in MIS, such as noise, blur, and low contrast. CNN architecture has several different backbones that are used in MIS, such as VGG (2014, [4]), ResNet (2015, [5]), DenseNet, DeepLabv3, MobileNets (2017, [6][7][8]), EfficientNet (2019, [9]).

- Recurrent Neural Networks (RNN) (1989, [10]) Barak A. Pearlmutter has explored various approaches to constructing the foundational concept of continuous time-recurrent networks. The Recurrent Neural Network (RNN) aims to mitigate the weight gradient to minimize the time-orbit error of the states in the continuous regression network. RNNs are particularly well-suited for continuous time domains such as signal, control, and speech processing. In medical image segmentation, RNN has been applied to model the time-dependence of image sequences (video). By leveraging the relationships between space-time information, RNNs can be combined with other architectures to improve the accuracy of image segmentation. RNNs can capture both local and global spatial features of the image by considering the context information [11].

TABLE I.    THE COMMON FRAMEWORK OF INTELLIGENT VISION SYSTEMS (IVS) FOR MEDICAL IMAGE SEGMENTATION (MIS)

| | Level of IVS | Explanation | Specific Solutions |
|---|---|---|---|
| 1 | **Backbone DNN Architectures** | Traditional segmentation methods, such as thresholding, edge detection, and region-based techniques, have limitations when dealing with noisy, fuzzy, and low-contrast medical images. Convolutional Neural Networks (CNNs) alleviate these drawbacks. Additionally, Graph Neural Networks (GNNs) and Transformers address the limitations of convolution kernels with regards to locality and invariance | **Background Networks Architectures:** *CNN* (1982, [3]), *RNN* (1989, [10]), *GNN* (2009, [12]), *GAN* (2014, [14]), *Transformer* (2017, [16]). **Backbone DNN:** VGG (2014, [4]), ResNet (2015, [5]), DenseNet, DeepLabv3, MobileNets (2017, [6][7][8]), EfficientNet (2019, [9]). **Some additional modules:** Inception (2015, [19]), Dilation convolution (2016, [21]), ASPP (2017, [7]), Attention (2015, [29]), Squeeze-and-excitation (2018, [33]), Residual (2015, [5]), Dense (2017, [6]), Residual Dense (2018, [37]). |
| 2 | **Specific DNN Architectures for MIS** | UNet is a widely used method in medical image segmentation. Its variants and hybrid network architectures have been developed to achieve higher accuracy in segmentation | *Unet* (2015, [38]), *V-net*, 3D U-Net (2016, [39], [40]), Mask R-CNN (2017, [41]), *U-Net++* (2018, [42]), UNet 3+, DRU-Net, DoubleU-net (2020, [43] – [45]), *TransUNet*, Swin Transformer (2021, [46], [47]), UNETR++, LE-UDA, *TransUNet+* (2022, [48], [49], [50]), SEP (2022, [51]), ESFPNet (2022, [52]) |
| 3 | **Applications of MIS** | Medical image segmentation has evolved from single-organ segmentation to multi-organ segmentation and from organ segmentation to lesion segmentation. This has allowed for a more comprehensive and effective diagnosis of diseases, thereby improving human health care. | *Kidney Tumor Segmentation* (F.Isensee et al., 2019, [53]), Brain Tumor Segmentation (S. Li et al. 2021, [54]), COVID-19 infection localization and severity grading from chest X-ray images (A. M. Tahir et al, 2021, [2][1]), *Abdominal Multi-Organ Segmentation* (F.Isensee, et al., 2022, [55]), and so on. |

- Graph Neural Network (GNN) (2009, [12]) Scarselli et al. introduced the Graph Neural Network (GNN) architecture, which expands upon traditional neural network methods for processing data represented in graph regions. Geometric Deep Learning, also known as GNN, is a nascent field of study that extends deep neural modeling to non-Euclidean domains. The structures of medical images often have irregular and unordered patterns, making it challenging to represent them as matrices for CNNs. As a result, graph-based representations are becoming increasingly popular in MIS [13].

- Generative Adversarial Networks (GAN) (2014, [14]) The Generative Adversarial Network (GAN) architecture was introduced by Goodfellow et al. based on game theory. In this architecture, two players, a generator and a discriminator, play against each other to minimize their respective costs. The discriminator's cost encourages it to correctly classify data as real or fake, while the generator's cost encourages it to generate the most realistic fake samples that the discriminator finds difficult to distinguish. In medical image segmentation, GANs can be used to create synthetic medical images and their corresponding segmented masks, leading to improved segmentation accuracy thanks to GAN's powerful generation ability and ability to capture the data distribution [15][14].

- Transformer (2017, [16]): The Transformer architecture was originally designed for natural language processing (NLP) tasks, where it achieved remarkable improvements. Its success in NLP has drawn the attention of the computer vision community. The Transformer enables parallel processing of input sequences while supporting long-term dependencies between sequence elements, thus overcoming the explicit long-term dependency limitations of the Unet model [17]. Transformers, unlike CNNs, are designed with less inductive bias and can fit into any data structure as easily as established functions. Their fundamental structure also demonstrates great scalability, making them suitable for networks with high capacity and large data sets. This allows for multi-modal processing, including images, videos, text, and audio, using the same processing blocks. For medical imaging, organs that are frequently spread across a large receptive field can be efficiently encoded by modeling relationships between distant pixels. Therefore, the ability of Transformers to model the global context is crucial for accurate medical image segmentation, such as lung segmentation. Additionally, medical images are often blurred, noisy, and have low contrast, such as in ultrasound scans. Understanding the overall context between pixels against the background can help models avoid mis-segmentation [18].

*2) SOTA modules for enhancing segmentation efficiency*

- For aggregating features at multiple scales:

  - The Inception module (2015, [19]) which concatenates multiple parallel convolutional filter banks with varying kernel sizes to extract features at multiple scales. Example of the application of the Inception module in medical image segmentation can be found in [20].
  - The dilation convolution (or "atrous convolution") kernel, introduced in 2016 [21], increases the size of the kernel, and the corresponding receptive field without significantly increasing the number of pixels processed. This results in improved speed and accuracy. Example of the use of the dilation module in medical image segmentation networks can be found in [22].
  - The Atrous Spatial Pyramid Pooling (ASPP, [7]) module, introduced in 2017, utilizes dilated (or atrous) convolution to gather information at multiple scales. This helps to preserve local features while capturing multi-scale contextual information, leading to improved segmentation efficiency. ASPP module is applied in medical image segmentation such as retinal segmentation [23], segmentation of abdominal organs from CT images [24], SAR-U-Net liver segmentation from CT images [25], U-Net-ASPP segmented COVID-19 [26], localized skin lesions [27]; spinal segments [28]

- Focusing on important features:

  - The attention mechanism (2015, [29]) focuses on spatially significant features. This mechanism is commonly applied in medical image segmentation problems such as [30] - [32]
  - Squeeze-and-excitation block (2018, [33]) focus on features based on channel-weighted adjustment. Medical image segmentation problems that apply this block like [34] - [36].

- Connecting to the previous layers and solving the vanishing gradient problem:

  - Residual block (2015, [5]), which adds the previous layer outputs to feature maps learned from the current layer;

- Dense block (2017, [6]), which connects the outputs of all previous layers to the feature maps learned by the current layer;
- Residual Dense block (2018, [37]), which allows full using the local and global features.

Level 2: Specific DNN Architecture for MIS.

Background network architectures at level 1 such as CNN, GNN, transformer, and so on can be used to implement various tasks such as detection, recognition, classification, and segmentation. This paper focuses on DNN architectures for **image segmentation.** TABLE II. compares three specific networks for the medical imaging segment, UNet (2015, [38]), UNet++ (2018, [42]), and TransUNet+ (2022, [50]) in terms of network architecture, pros, cons, and performance.

From the comparison results, it can be observed that network architectures are continuously improving in terms of performance. The current trend is the use of hybrid networks, such as TransUNet+ [50], to meet the increasing demand in healthcare systems.

TABLE II. COMPARES THREE SPECIFIC NETWORKS FOR THE MEDICAL IMAGING SEGMENT

| | Unet (2015, [38]) | Unet++ (2018, [42]) | TransUNet+ (2022, [50]) |
|---|---|---|---|
| **Architectures** | 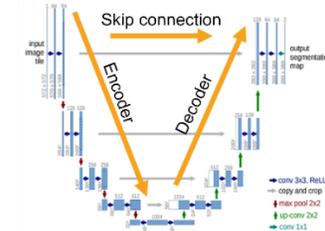 Unet includes **encoder** to extract features, **decoder** synthesize extracted features to segmentation results and **skip connection** copies low-resolution (encoder) to high-resolution (decoder) feature maps. | 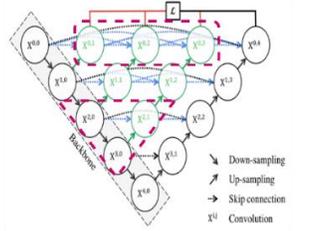 Uses dense blocks to **Re-designed skip pathways** and use **Deep supervision** | 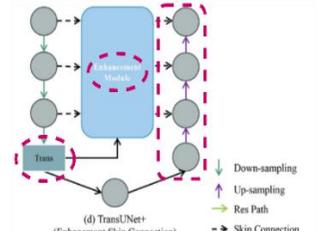 **Encoder**: CNN and Trans; **decoder:** original decoder and the enhanced features; **skip connection:** enhancement module. |
| **Pros** | - Train with fewer annotated images (at most 35 annotated images)<br>- Fast training time in 2015 (On a NVidia Titan GPU (6 GB), segmenting a 512x512 image takes less than a second.).<br>- Easily applied to more tasks. | - Added redesigned skip pathways and deep supervision for *more precise segmentation, model pruning, and increased speed.* | - Combining Transformer (*global self-attention mechanisms*) and CNN (enhance finer details by recovering *localized spatial information*), Redesigning the skip connection to *enhance features and improve the focus on the key patches*. In the decoder, cascaded up-sampler contains an up-sampling layer and a linear layer.<br>- Performance in small organ segmentation. |
| **Cons** | - Limits on network depth and skip connection<br>- Limitations on long-range information extraction | - Reduces the robustness of feature representation and increases the number of parameters. | - High computational cost and memory usage |
| **Performances** | 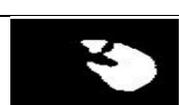 IoU: 76.62 (**LiTS Challenge**) | 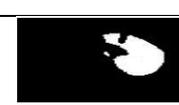 **IoU: 82.90** (**LiTS Challenge**)<br>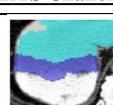 DSC 76.09 (**Synapse multi-organ CT dataset**) | 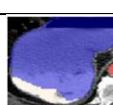 **DSC: 81.57** (**Synapse multi-organ CT dataset**) |

The paper not only compares three of the most specific network architectures for MIS, but it also tracks the latest MIS rankings on paperswithcode. According to the latest updates, the two top models, SEP (2022, [51]) and ESFPNet-L (2022, [52]), are at the forefront on two datasets, Kvasir-SEG and CVC-ClinicDB, respectively.

Spatially Exclusive Pasting (SEP) (2022, [51]) is an innovative data augmentation technique designed to tackle the issue of data scarcity in polyp segmentation, an important task in the diagnosis of intestinal diseases such as tumors and precancerous lesions.

Fig. 2 illustrates the procedure of SEP technique. The core concept of this technique is to copy the polyp region and paste it to other locations in order to generate a large number of new images. The augmentation process is divided into three modules: (1) a Potential Map Generation Module that generates a potential value for each coordinate, (2) a Pasting Module, and (3) an Update Module that updates the potential values for each coordinate.

According to the latest statistics on the rankings of paperswithcode.com (Fig. 3), SEP has achieved the highest mean Dice score (0.941) surpassing UNet (0.818), UNet++ (0.821), and FCBFormer (0.939).

The limitation of SEP is that it is only applicable to a limited number of data sources and is specifically designed for the task of polyp segmentation.

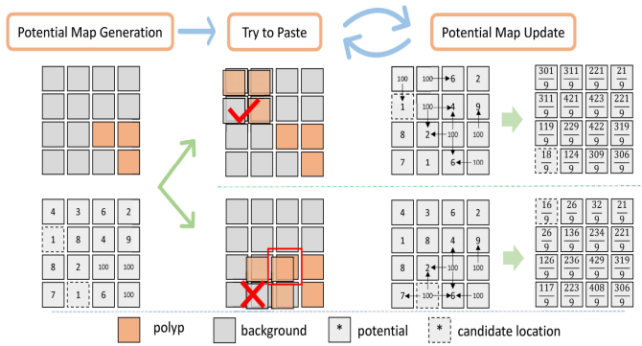

Fig. 2.    The process of SEP [51].

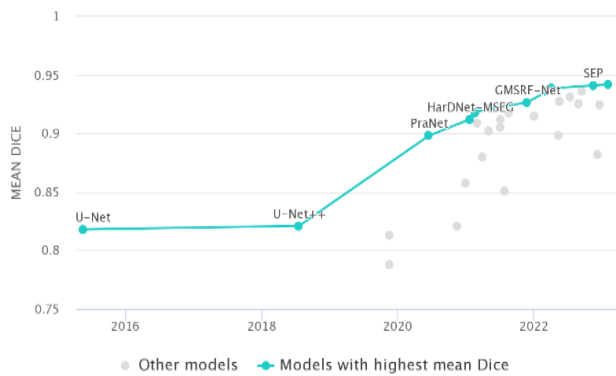

Fig. 3.    Leaderboard the models with the highest mean Dice on the Kvasir-SEG.

ESFPNet (2022, [52]) is a deep learning architecture designed for real-time accurate segmentation and robust detection of bronchial lesions in autofluorescent bronchoscopy (AFB) video streams. Fig. 4 depicts the architecture of ESFPNet, which consists of a pre-trained Mix Transformer (MiT) encoder that leverages the encoder structure and an efficient Intelligent Phased Feature Pyramid (ESFP) decoder structure. ESFPNet-L produces superior results, with a mean Dice score of 0.949, compared to other recent architectures such as DuAT (0.948) and ColonFormer (0.947). Fig. 5 displays the rankings of the models with the highest mean Dice scores on the CVC-ClinicDB dataset. Additionally, with a processing speed of 27 frames per second, ESFPNet provides clinicians with a useful tool for confidently segmenting and detecting lesions in real-time during direct airway bronchoscopy. However, one drawback of ESFPNet is the high cost and difficulty in acquiring more live human video data.

### Level 3: Applications of MIS

Level 3 is dedicated to the investigation of specific applications designed to perform medical image segmentation. This inherits and develops the background and backbone of the Deep Neural Network (DNN) architectures from Level 1 with a specific DNN architecture designed for image segmentation processing at Level 2.

According to the estimate of cancer cases in the United States in 2022, the most common types of cancer were prostate (268,490 cases in men) and breast (287,850 cases in women). The second most common cancers were lung and bronchus (117,910 cases in men and 118,830 cases in women) [56].

TABLE III. lists the number of papers with codes corresponding to each task, based on the latest statistics (as of November 2022) available on paperswithcode.com in the "Browse SoTA > Medical > Medical Image Segmentation" section.

Furthermore, the paper considers papers that have won first place in challenges at the International Conference on Medical Image Computing and Computer-Assisted Intervention (MICCAI)[1] in the past three years (2019 - 2022). TABLE IV. summarizes these challenges [57]

According to the research results, it is evident that state-of-the-art (SOTA) applications are focusing on improving segmentation performance, from single organ segmentation to multi-organ segmentation (abdomen, brain, genital organs, etc.), and from organ segmentation to tumor and infection area segmentation, to provide comprehensive care for human health.

### B.  Data

*1)  Medical image modalities:* Data is a crucial component in the learning process. Medical imaging is a method and approach to create visual images of the interior of the body using non-invasive technologies [58]. It is used to aid in the diagnosis or treatment of various diseases. Some common

---

medical imaging techniques include X-rays, computed tomography (CT) scans, magnetic resonance imaging (MRI), ultrasound (US), positron emission tomography (PET), and single-photon emission computed tomography (SPECT), among others [59]. This paper explores the four most commonly used medical imaging modalities—X-ray, CT, MRI, and Ultrasound. TABLE V. compares these four imaging modalities in terms of their advantages, disadvantages, SOTA applications, and health effects [60].

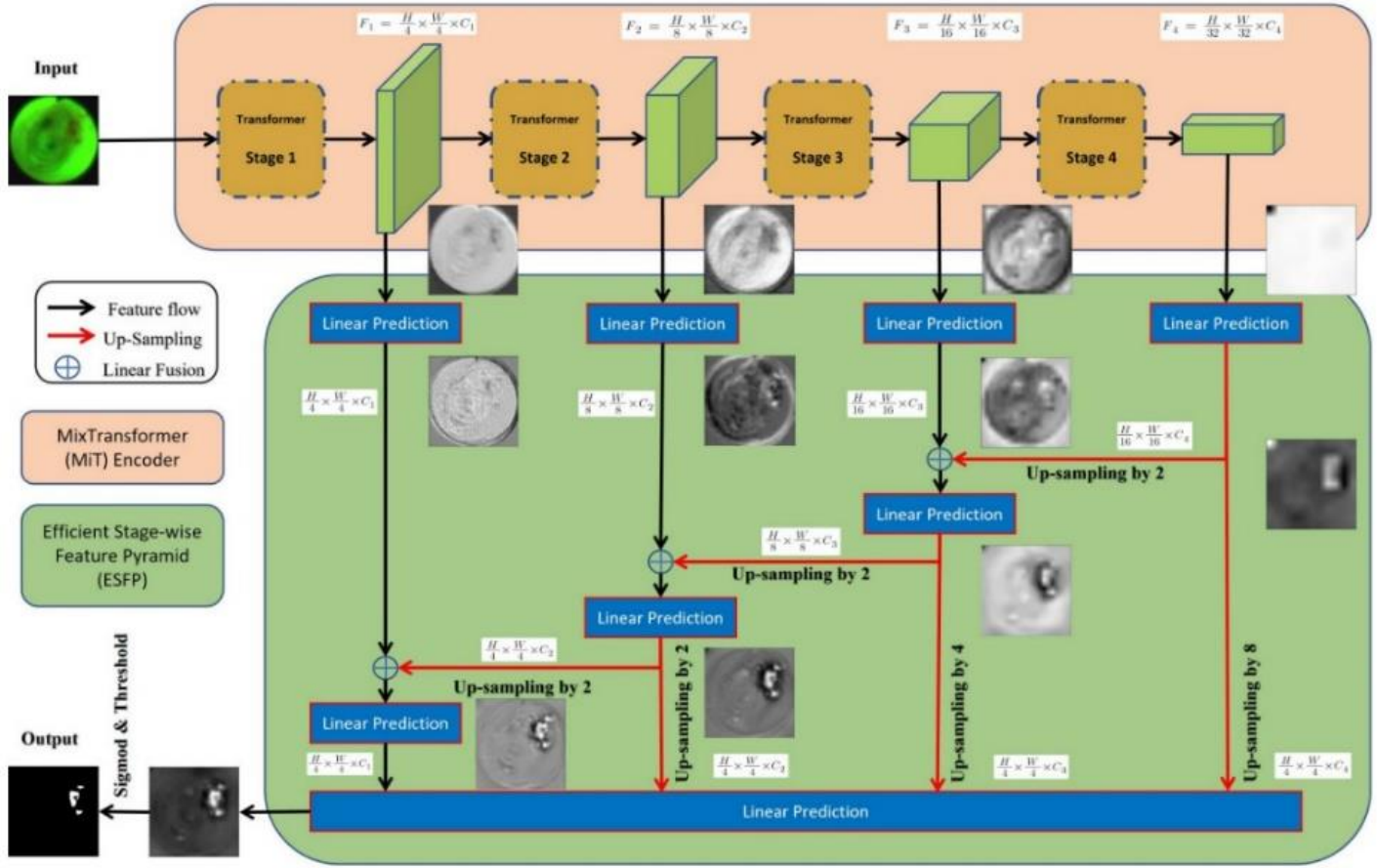

Fig. 4. ESFPNet architecture [52].

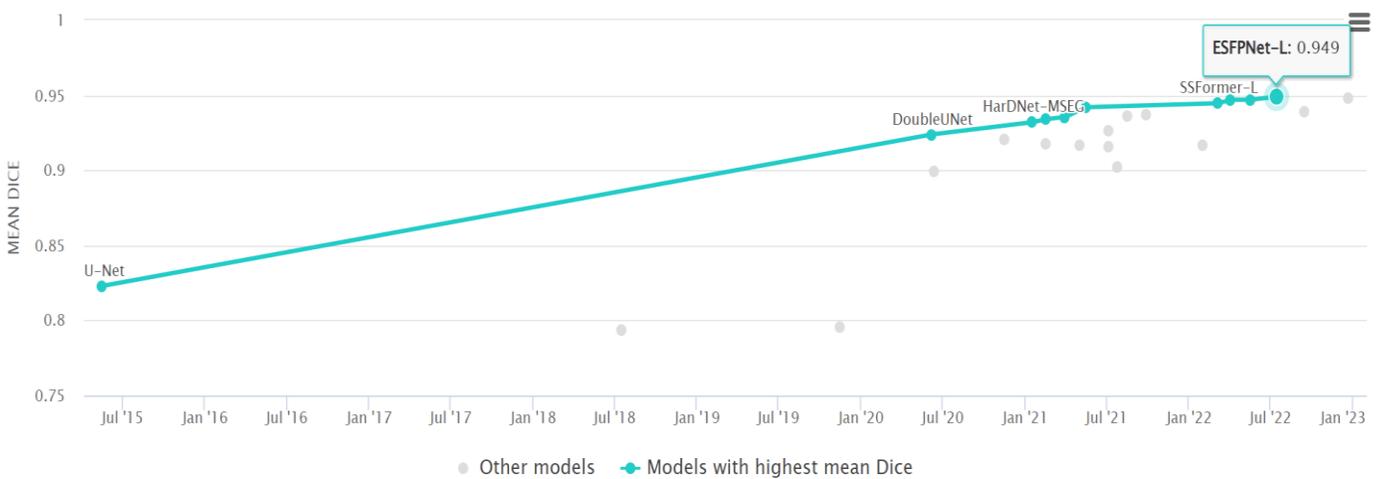

Fig. 5. Leaderboard the models with the highest mean Dice on the CVC-ClinicDB.

TABLE III.    STATISTICS OF NUMBER OF PAPERS WITH CODE ON PAPERSWITHCODE.COM[2]

| | Segmentation tasks | Benchmarks | Papers with code | | Segmentation tasks | Benchmarks | Papers with code |
|---|---|---|---|---|---|---|---|
| 1. | Medical Image | 104 | 407 | 17. | Video Polyp | 4 | 11 |
| 2. | Lesion | 8 | 142 | 18. | COVID-19 Image | | 10 |
| 3. | Brain Tumor | 9 | 94 | 19. | Lung Nodule | 5 | 8 |
| 4. | Brain | 1 | 51 | 20. | Nuclear | 1 | 8 |
| 5. | Cell | 8 | 41 | 21. | Skin Cancer | 2 | 8 |
| 6. | Skin Lesion | 2 | 39 | 22. | Electron Microscopy Image | 3 | 7 |
| 7. | Retinal Vessel | 4 | 36 | 23. | Infant Brain Mri | 1 | 5 |
| 8. | MRI | | 32 | 24. | Brain Lesion From Mri | | 5 |
| 9. | 3D Medical Image | 3 | 28 | 25. | Ischemic Stroke Lesion | | 4 |
| 10. | Cardiac | | 26 | 26. | Automatic Liver And Tumor | | 3 |
| 11. | Liver | 1 | 23 | 27. | Placenta | | 3 |
| 12. | Semi-supervised MIS | 2 | 17 | 28. | Acute Stroke Lesion | | 1 |
| 13. | Brain Image Segmentation | 6 | 14 | 29. | Cerebrovascular Network | | 1 |
| 14. | Volumetric MIS | 1 | 12 | 30. | Automated Pancreas | | 1 |
| 15. | Pancreas | 2 | 12 | 31. | Semantic Segmentation of Orthoimagery | | 1 |
| 16. | Iris | 3 | 12 | 32. | Pulmonary Vessel | | 1 |

TABLE IV.    SUMMARY OF CHALLENGES AT THE INTERNATIONAL CONFERENCE MICCAI IN THE LAST THREE YEARS (2019 - 2022)[3] [57]

| | Challenges | First Author | Title |
|---|---|---|---|
| 1. | 2022 MICCAI: Multi-Modality Abdominal Multi-Organ Segmentation Challenge (AMOS22) (Results) | Fabian Isensee, Constantin Ulrich and Tassilo Wald | Extending nnU-Net is all you need (paper) (code) |
| 2. | 2021 ISBI: MitoEM Challenge: Large-scale 3D Mitochondria Instance Segmentation (MitoEM) (Results) | Mingxing Li | Advanced Deep Networks for 3D Mitochondria Instance Segmentation (paper) (code) |
| 3. | 2021 MICCAI: Fast and Low GPU memory Abdominal oRgan sEgmentation (FLARE) (Results) | Fan Zhang | Efficient Context-Aware Network for Abdominal Multi-organ Segmentation (paper) (code) |
| 4. | 2021 MICCAI: Kidney Tumor Segmentation Challenge (KiTS) (Results) | Zhaozhong Chen | A Coarse-to-fine Framework for The 2021 Kidney and Kidney Tumor Segmentation Challenge (paper) |
| 5. | 2020 MICCAI: Automatic Evaluation of Myocardial Infarction from Delayed-Enhancement Cardiac MRI (EMIDEC) | Yichi Zhang | Cascaded Convolutional Neural Network for Automatic Myocardial Infarction Segmentation from Delayed-Enhancement Cardiac MRI (arxiv) |
| 6. | 2019 MICCAI: Kidney Tumor Segmentation Challenge (KiTS19) | Fabian Isensee | Automated Design of Deep Learning Methods for Biomedical Image Segmentation (arxiv). |

---

[2] https://paperswithcode.com/area/medical/medical-image-segmentation
[3] https://github.com/JunMa11/SOTA-MedSeg

TABLE V.　　Comparison between the Medical Imaging Modalities

| | X-ray | CT | MRI | US |
|---|---|---|---|---|
| **Advantages** | Low cost<br>Fast imaging time. | Quick imaging<br>Excellent spatial resolution<br>It is possible to combine it with angiographic techniques. | There is no ionizing radiation.<br>Exceptional spatial resolution<br>Outstanding soft tissue contrast<br>Dynamic angiographic imaging. | Low cost and real-time nature<br>Fast imaging time<br>No ionizing radiation, good spatial resolution<br>More prevalent, portability. |
| **Disadvantages** | Ionizing radiation<br>Low sensitivity. | Ionizing radiation<br>Low sensitivity<br>Limited soft tissue contrast. | High cost<br>Long imaging time<br>Contraindications in some patients | Operator dependent<br>The difficulty of distinguishing imaging structures between tissue and gas<br>Noise, shadow, speckle, low contrast, and blurred edges<br>Limited penetration/sensitivity |
| **Applications of MIS** | Bone [61], Lung [2], [62], Caries lesion [63] | Lung [42], [64] - [66], Proximal femur segmentation [67], Kidney tumor segmentation [68], tooth and alveolar bone segmentation [69] | Brain [66], [70] - [72], Retinal Vessel [73] - [75], Cardiac [66], [76] – [78], prostate [66], [39], [78], Osteosarcoma [79] | Breast [80], kidney [81], prostate [82], multi-organs [83] |
| **Health effects** | Biological effect, need protection against unnecessary does. | High radiation, dangerous to health. | Less harmful effects, better for the fetus. | Safe, painless, non-invasive and non-ionized. |

Some comments on medical imaging modalities:

- Regarding the segmentation problem: Due to challenges such as noise, shadow, speckling, low contrast, and blurred edges, US images pose more difficulties in segmentation compared to MRI and CT images

- Regarding health effects: X-ray and CT imaging have various negative impacts on human health.

- Regarding segmentation applications: Multiple medical imaging modalities can be utilized for various health care applications to provide comprehensive health care for individuals

*2) MIS in different dimensionality of medical images:* This section provides an overview of 2D, 3D, and 4D data types in medical imaging, as well as image segmentation issues in these three data types.

- Structured 2D images have a defined height and width and exist in a flat space. The most common type of medical image in this category is X-rays. This paper focuses on examining segmentation models, such as 2D CNNs, original UNets, and their variants, that are applicable to 2D image.

- Structured 3D images have a defined height, width, and depth, and can be considered as a collection of stacked 2D frames. This type of medical imaging, which utilizes spatial relationships, is commonly used in modalities such as CT and MRI scans. 3D imaging is widely used in clinical practice due to its ability to provide rich information about the imaged regions, which aid in the visualization and quantification of various tissues and organs. However, manual segmentation of 3D images is challenging and time-consuming, highlighting the importance of automated computer-aided segmentation models. or 3D medical image segmentation, CNN models often employ a 3D kernel to extract spatial features or utilize GNN or Transformer. Some of SOTA models in this field include 3D U-Net [40], V-Net [39], nnU-Net [84], HighRes3dNet [85], 3D-Res-Unet [86], DenseVNet [87], UNETR [88], SegResNet [89], Point-Unet [90], and others.

- Structured 4D images, which are comprised of height, width, depth, and temporal dimensions, are commonly referred to as dynamic volumes and represent moving data in real-time. In the medical field, 4D images are often used to measure various parameters such as heart rate [91], lung breathing [92], blood flow rate, fetal movement, and more. In the field of medicine, 4D imaging techniques include dynamic volume CT, 4D CT, MRI, and ultrasound. The segmentation of 4D medical images has great potential in uncovering disease progression and monitoring disease trajectory [93]. However, traditional image segmentation can be complex and costly without the aid of AI models. Commonly used models in 4D image segmentation include "balloon" models and deformable (time-varying) models, such as, LSTM, FCSLSTM [93], and XCAT [94].

Based on the characteristics of each medical imaging modality, as well as the statistics from the MICCAI 2023 and 2022 challenges, and the latest papers on paperswithcode, it appears that 3D medical imaging techniques, such as MRI and CT, are currently the most popular. The advantages of 3D medical images include the utilization of spatial relationships, the ability to obtain more information from image regions than can be obtained from 2D images, and a lower cost for processing and acquisition compared to 4D images.

*3) Medical image dataset:* Data is a crucial aspect of learning, and medical image segmentation datasets often contain sensitive information, making them highly privacy-sensitive. However, in order to evaluate the performance of image segmentation methods, the datasets must be made publicly available. To accomplish this, organizers gather and anonymize medical imaging datasets, and host various challenges to advance the field of medical imaging. There are several medical databases available for free use in research, including:

- World Health Organization (WHO): WHO is a United Nations agency that connects nations, partners, and people to promote health and ensure the safety of the world. Established in 1948, WHO serves the most vulnerable populations so that everyone, everywhere, can attain the best possible health. The World Health Data Hub of the WHO is a comprehensive digital platform for global health data that provides efficient solutions for collecting, storing, analyzing, and sharing timely, reliable, and actionable data. The WHO is responsible for managing and preserving a vast array of data collections related to global health and well-being, as mandated by its Member States.

- Medical ImageNet is a large-scale resource for machine learning based on medical images. The Stanford Center for Artificial Intelligence in Medicine and Imaging (AIMI Center) was established in 2018 with the primary objective of using AI to address clinically significant medical issues. The Stanford Medical ImageNet is a petabyte-scale searchable repository of annotated and de-identified clinical images (radiology and pathology) that are linked to genomic data and electronic medical records, providing a platform for the speedy development of computer vision systems.

- Kaggle: Kaggle offers a customizable Jupyter Notebook environment with no setup required. It provides free access to GPUs and a vast repository of community-submitted data and code. Within Kaggle, you will find all the necessary code and data to complete your data science projects. With over 50,000 public datasets and 400,000 public notebooks available, you can quickly complete any analysis. Currently, Kaggle has 930 medical datasets, including 210 medical image datasets.

- Paperswithcode: The mission of Papers with Code is to create a free and open resource for machine learning research, including papers, code, datasets, methods, and benchmarks. It lists 2029 results for medical image datasets. Simpson, Amber L., et al. released ten datasets for medical image segmentation for various tasks under the Medical Segmentation Decathlon.

- mridata.org is an open platform for researchers to share raw magnetic resonance imaging (MRI) datasets. The website was created through a collaboration between Professor Michael Lustig's group at UC Berkeley and Professor Shreyas Vasanawala's group at Stanford's Lucile Packard Children's Hospital. The datasets available on the website can be used for a variety of purposes.

- Medical image datasets for segmentation are collected by researchers from various sources, including scientific journals and conferences, and collaborations between health organizations and partnerships. These datasets can be found by searching Google Scholar and Google using keywords such as 'medical segmentation', 'medical image datasets for segmentation', etc.

*4) Taxonomy of data-driven learning paradigms:* Data-driven learning paradigms can be classified into four categories: supervised learning, unsupervised learning, semi-supervised learning, and weakly supervised learning.

- Supervised learning: The first type of machine learning that humans introduced was supervised learning. In this method, the learning model is trained by providing labeled data, where both the input data and the corresponding output labels are given. The model then uses this information to make predictions on new, unseen data based on what it has learned from the training data. Learning from a forest to predict a tree is an example of supervised learning. The first model to utilize this method was the perceptron. Most ML algorithms employ supervised learning. In the medical field, supervised learning is commonly used to estimate risk. With risk modeling, the computer not only replicates a doctor's expertise, but it can also uncover novel relationships that might not be noticeable to humans [95].

- Unsupervised learning: However, labeling data is often difficult, expensive, and time-consuming, requiring the assistance of specialists. Furthermore, as network architecture advances, models can learn from unlabeled data. This is known as unsupervised learning. The Deep Belief Network (DBN) was proposed by Hinton in 2006, which uses this unsupervised mechanism. Unsupervised learning is based on data structures for performing clustering and data classification. However, since the data is not labeled, the learning challenge will increase. The usefulness of the patterns identified through unsupervised learning needs to be assessed either by human inspection or through additional supervised learning tasks [96]. Studies using unsupervised learning for Medical Image Segmentation (MIS) include the following: breast cancer segmentation on MRI [97]; COVID segmentation on CT lung tissue [98]; brain segmentation on 3D MRI [99]; multi-modality segmentation, including 2D hand x-ray, 3D abdominal magnetic resonance (MR) image, and 3D cardiovascular MR images [100], as well as cardiac substructure segmentation and abdominal multi-organ segmentation between MRI and CT images [101], etc.

- Semi-supervised learning combines supervised and unsupervised learning paradigms to solve the problem of training an accurate classifier with less human effort and time. This framework leverages both a limited amount of labeled data and a large amount of unlabeled (undiagnosed) data to achieve this goal [102]. Studies using semi-supervised learning in the field of MIS can be found in the range of studies [103] to [105].

- Weakly supervised learning paradigms are used when raw data is not fully processed or processed inaccurately. The goal of weakly supervised learning is similar to that of supervised learning, but instead of using a carefully labeled and processed training set, weak supervision is provided through one or more weakly annotated examples. These examples may come from community sources, be the output of rule heuristics, the results of remote monitoring, or the output of other classifiers [106]. Numerous studies have employed weakly supervised learning for the segmentation of medical images, as demonstrated in researches [107] to [109].

One thing to remember about medical imaging data is that the amount of annotated (labeled) data is very small. TABLE VI. TABLE VI. compares the four paradigms discussed above in terms of labeling costs.

TABLE VI. COMPARISON OF LABELING COSTS OF DATA-DRIVEN LEARNING PARADIGMS

| Methods | Data | Labeling Cost |
|---|---|---|
| Supervised Learning | Labeled data | High |
| Unsupervised Learning | Unlabeled data | No |
| Semi-supervised Learning | Labeled data + unlabeled data | Medium |
| Weakly supervised Learning | Labeled data (small) + noise data (incomplete and inaccurate labels) + unlabeled data | Low |

The accuracy of data-driven learning models is directly proportional to the amount of annotated data and inversely proportional to the cost of labeling. To improve the accuracy of learning models while reducing the cost of labeling, two solutions that have been proposed are transfer learning and active learning.

Traditional machine learning methods assume that the training and test data come from the same domain, with the same input feature space and data distribution characteristics. However, this assumption is not always true in real-world machine learning scenario. In some cases, collecting and annotating training data can be expensive or difficult. Transfer learning addresses this issue by training models on labeled data from domains where data collection and annotation are easier. The knowledge gained from the training data is then transferred to the test data domain [110]. Many models today are pre-trained on the ImageNet dataset [111]. Transfer learning has been applied in the field of medical image analysis in works such as [112] and [113].

While the ImageNet dataset is large and well-labeled, medical image data has distinctive characteristics that set it apart from natural images in datasets like ImageNet. Active learning is a method of selecting a subset of data from a larger dataset, such as a data lake [114] for annotation. The goal of this method is to increase the amount of annotated data, reduce data annotation costs, and improve model performance.

This paradigm has been used in papers ranging from [115] to [118]

### C. Loss Functions

*1) Common loss functions:* The image segmentation labels each pixel with the corresponding class. Therefore, it is necessary to use a mechanism to calculate the loss weight for each pixel. The most commonly used loss functions in segmentation are cross entropy and its variants.

- Cross-Entropy loss is a fundamental function in medical image segmentation. It is derived from the Kullback-Leibler (KL) divergence, which measures the difference between two probability distributions, such as those provided by the training set. The minimum KL divergence is equivalent to the minimum Cross-Entropy. The Cross-Entropy is defined as follows:

$$L_{CE} = -\frac{1}{N}\sum_{i=1}^{N}\sum_{c=1}^{C} gt_i^c \log ms_i^s \qquad (1)$$

where $gt_i^c$ is a binary indicator representing whether class label c is the correct classification for pixel i, and $ms_i^s$ is the corresponding predicted probability.

- A variant of the Cross-Entropy loss is the Weighted Cross-Entropy loss (WCE). This loss function takes into account class imbalance by assigning weights to different classes. Another emerging variant of the Cross-Entropy loss is the Focal Loss, which adjusts the weights of well-classified training samples to reduce their impact.

$$L_{WCE} = -\frac{1}{N}\sum_{c=1}^{C}\sum_{i=1}^{N} w_c \, gt_i^c \log ms_i^s \qquad (2)$$

where $w_c$ is the class weight assigned to penalize majority classes, $w_c$ is typically set inversely proportional to the frequency of each class in the training set. In the experiments to follow, we will set the class weight $w_c$ to be the reciprocal of class frequency in the training set.

Aside from cross-entropy, other standard loss functions used in image segmentation are the **Dice loss**, and the **Intersection over union (IoU) loss** - which is derived from the Jaccard index and measures the ratio of sample intersection to its union. Dice loss and IoU loss are often used to improve the corresponding evaluation metrics.

$$L_{Dice} = 1 - \frac{2\sum_{c=1}^{C}\sum_{i=1}^{N} gt_i^c ms_i^s}{\sum_{c=1}^{C}\sum_{i=1}^{N} gt_i^c + \sum_{c=1}^{C}\sum_{i=1}^{N} ms_i^s} \qquad (3)$$

$$L_{IoU} = 1 - \frac{\sum_{c=1}^{C}\sum_{i=1}^{N} gt_i^c ms_i^s}{\sum_{c=1}^{C}\sum_{i=1}^{N}(gt_i^c + ms_i^s - gt_i^c ms_i^s)} \qquad (4)$$

*2) Hybrid loss functions:* Jun Ma et al. presented an overview of loss functions in MIS sorted according to the following classification system: distribution-based, region-based, boundary-based, and association-based. At the same time, the authors also found the relationship (connection) between different loss functions (see Fig. 6), as well as the specific use case of loss functions in different applications of MIS [119]. These loss functions are installed on GitHub [4]

**Loss function relationships:** As shown in Fig. 6 there are strong connections between loss functions.

---

[4] https://github.com/JunMa11/SegLoss.

- Most distribution-based and region-based loss functions are variants of Cross entropy and Dice loss.

- Although boundary-based losses are designed to minimize the distance between two boundaries, they share some similarities with Dice loss because both are calculated using region-based methods.

- A compound loss is a combination of multiple loss functions.

**Recommendations for selecting loss functions:** It is impossible to determine which loss function is the best. The data balance can be used to select the appropriate loss function.

- Mild imbalance issues are well handled by Dice loss or General Dice loss (GD)

- For highly imbalanced segmentation tasks, the compound loss functions are more commonly used.

The Combo Loss function proposed by Saeid Asgari Taghanaki et al. [120] aims to improve multi-organ segmentation performance in cases where the input and output are imbalanced. This loss function has been shown to achieve higher Dice scores and reduce false negatives and false positives, and can be applied to 3D U-Net, 3D V-Net, and 3D Seg-Net architectures.

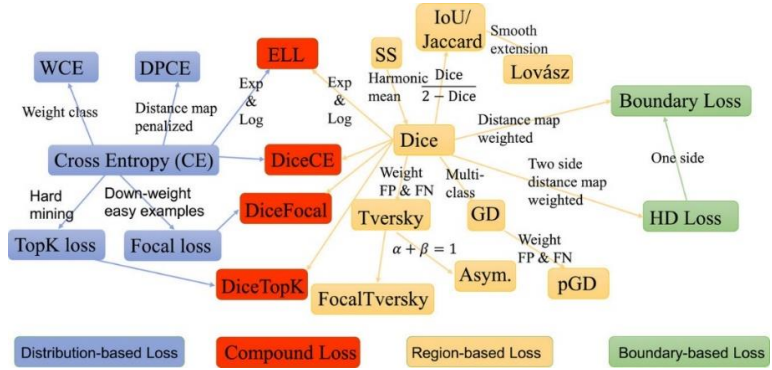

Fig. 6. An overview of 20 loss functions for MIS and their relationships [119].

### D. Evaluation Metrics

Segmentation performance evaluation involves comparing the similarity between a manually generated and a DNN-generated segmentation. Many metrics are used for evaluation, but some of the most commonly used ones in medical imaging are presented below.

In MIS, the regions of interest (ROI) are often quite small compared to the overall image. This results in an imbalanced distribution of pixels between different classes. To address this issue, two common metrics used in MIS are the Dice Similarity Coefficient (DSC), also known as the F1 score, and the Intersection over Union (IoU), also known as the Jaccard index.

Explanation of some acronyms: GT: Ground Truth; MS: Machine Segmentation; TP: True positive; TN: True negative; FP: False positive; FN: False negative; these terms are illustrated in Fig. 7, which has been modified from source [121].

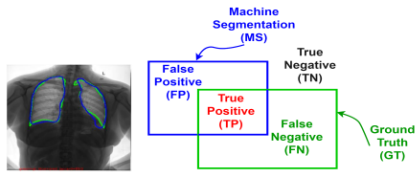

Fig. 7. Illustrate GT, MS, TP, TN, FP, and FN modified from the source [121].

*1) Jaccard index (JAC - IoU) evaluates the overlap between GT and MS regions.*

*2) Dice similarity coefficient (DSC - F1 score) measures the similarity between GT and MS regions.*

$$JAC = IoU = \frac{|GT \cap MS|}{|GT \cup MS|}$$
$$= \frac{TP}{TP + FP + FN} \quad (5)$$

$$DSC = F1\ Score = \frac{2 * \; \boxed{}}{\boxed{} + \boxed{}} = \frac{2|GT \cap MS|}{|GT| + |MS|}$$
$$= \frac{2TP}{2TP + FP + FN} = \frac{2JAC}{1 + JAC} \quad (6)$$

The following are other metrics that are less commonly used in the evaluation of medical image segmentation due to their sensitivity to the size of the segment, meaning they penalize errors in small segments more heavily than in larger segments.

*3) The Hausdorff distance (HD) between two GT and MS regions is defined as "the maximum of all minimum distances" (see Fig. 8) [122].*

$$HD(GT, MS) = max\{d(gt, ms)\}\} \quad (7)$$

where gt and ms represent the pixels of regions GT and MS respectively, and d (gt, ms) is any metric between these pixels; for the sake of simplicity, take d (gt, ms) is taken as the Euclidean distance between gt and ms.

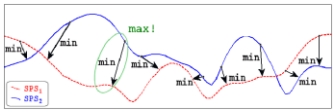

Fig. 8. Illustrate the Hausdorff distance [122].

### 4) Other evaluation metrics

- Sensitivity / Recall / True Positive Rate (TPR):

$$Sensitivity = Recall = TPR = \frac{TP}{TP+FN} \quad (8)$$

- Specificity or True Negative Rate (TNR):

$$Specificity = TNR = \frac{TN}{TN+FP} \quad (9)$$

- Accuracy (ACC):

$$Acc = \frac{TP+TN}{TP+FP+TN+FP} \quad (10)$$

- False Positive Rate (FPR)

$$FPR = \frac{FP}{FP+TN} = 1 - TNR \quad (11)$$

- False Negative Rate (FNR)

$$FNR = \frac{FN}{FN+TP} = 1 - TPR \quad (12)$$

- Precision or Positive Predictive Value (PPV):

$$Precision = PPV = \frac{TP}{TP+FP} \quad (13)$$

## III. EXPLAINABLE ARTIFICIAL INTELLIGENCE IN MEDICAL IMAGE SEGMENTATION (XAI IN MIS)

Recently, the use of AI in the medical field has led to impressive results, with machine learning models achieving over 99% accuracy. However, the practical application of these models remains limited due to their "black box" nature. XAI learns how models make decisions, investigates the inner workings of its layers, and provides visualizations of neural networks. Interpretability, which supports the reasoning behind a model's output, is crucial, particularly in precision medicine where experts require more in-depth information from a model beyond just a binary prediction to make an accurate diagnosis. In general, people are cautious about using techniques that are not transparent and can't be easily understood, due to the increasing emphasis on ethical AI, particularly in fields that have a direct impact on human lives [123]. To build trust among physicians, regulators, and patients, medical diagnostic systems must be transparent, easily understandable, and capable of providing explanations [123]. Moreover, by using XAI, model designers can uncover weaknesses in existing architecture and debug and fine-tune models to improve their performance.

If an image-based diagnostic system is enhanced with XAI, it will reach a level of Wisdom, which is higher than mere Intelligence. This is because, in addition to having high IQ, it also has high EQ or a sense of responsibility.

XAI can be divided into two categories: interpretability and explainability.

- "Interpretability" focuses on the underlying processes and events. It answers the "how" question by showing how the decision was made (based on the scoring criteria), but it does not explain "why" the criteria used are reasonable. The term "interpretability" addresses the "quantitative" aspects of the decision-making process.

- "Explainability" focuses on the inherent characteristics of events. It addresses the "why" question by explaining the reasoning behind the decision, rather than just the "how". The term "explainability" refers to the "qualitative" aspects of the decision-making process. Some commonly used methods for explainability include CAM [124], Grad-CAM [125], Grad-CAM++ [126], LIME [127], and SHAP [128]:

  - Bolei Zhou and colleagues introduced the Class Activation Maps (CAM) approach [124], which utilizes a global average pooling (GAP) layer at the end of the neural network. CAM provides an interpretation of the model and reveals the areas of the image that the network focuses on to make decisions. By producing heatmaps, CAM shows which regions of the image are most important for decision-making.

  - Building on the CAM approach, various variants have been developed for the XAI process, including the widely used Grad-CAM [125] and Grad-CAM++ [126].

  - Local Interpretable Model-agnostic Explanations (LIME) [127] were introduced by Marco Tulio Ribeiro and others as a method for creating an interpretable model that accurately reflects the classifier in a locally interpretable representation.

  - SHAP (SHapley Additive exPlanations) [128], developed by S. Lundberg and S.-I. Lee, is a method that assigns a significant value to each feature for a specific prediction. This approach is viewed as a comprehensive framework for interpreting predictions.

There are some XAI models that can only be interpretable, explainable or both.

### 1) Consider the linear regression model for disease diagnosis.

- It is considered "interpretable" because once the coefficients of the linear model are calculated, the new input can be used to determine how the predicted output was obtained. This process is clearly quantified.

- However, it is "unexplainable" because it does not provide an explanation for why there is a linear relationship between the independent and dependent variables, which is not clearly quantified.

- Both "interpretability" and "explainability" can be achieved if the statistical process includes a step for testing the linear hypothesis. This provides a way to determine the presence of a linear relationship and quantify the relationship between the independent and dependent variables, making the results both interpretable and explainable.

*2) Example of a model that can only be explained:* The lung infection segmentation model from X-ray images was trained using a Deep Convolutional Neural Network (DCNN) combined with Class Activation Maps (CAM) on data from two different scanners at two hospitals. It is possible to understand which areas of the lungs contribute to a positive outcome ("why"), but the reason why the sample test with the trained model at hospital 1 is positive but with the trained model at hospital 2 is negative is unclear ("how" to calculate it is not understood).

## IV. EARLY PREDICTION WITH MIS

This section presents the role and typical applications of MIS in the field of early prediction.

Tumors often start as small, hard-to-detect nodules, which can lead to a high mortality rate among patients. However, if detected early and treated promptly, it can increase the patient's chances of survival and decrease the cost of treatment. This is the motivation behind researchers performing early prediction problems. The segmentation results can be used to anticipate disease progression and offer appropriate treatment. Early prediction has a critical role in reducing patient risk, which holds great humanitarian significance. Furthermore, it helps to elevate the level of the system from intelligence to wisdom.

### A. Some Typical Applications

The Fuzzy C-means (FCM) intelligent segmentation algorithm [129] was developed for early detection of enlarged hematoma in patients with intracerebral hemorrhage (ICH) on CT images. The processing of cranial CT images using the FCM algorithm has high clinical value in predicting early hematoma in ICH patients.

Lung cancer is usually detected early when lesions appear in the bronchial epithelium of the airway wall. Autofluorescence bronchoscopy (AFB) [52] has been shown in recent studies to be particularly effective in detecting lesions, making it a potentially crucial approach for airway evaluation. Bronchoscopy is a commonly used method for detecting early-stage lung cancer. ESFPNet is a method for accurately segmenting and identifying bronchial lesions in AFB video streams.

There are also some other early prediction models based on segmentation results such as brain-related disease [130] - [132], evaluation of bone tumors [133], lung disease [134] - [136], breast cancer [137] - [138], tumor metastasis of ovarian cancer patients [139], stroke, ischemic coma [140] - [141], acute pancreatitis [142], and cancer radiation [143]- [144].

In addition, early segmentation is also applicable to cases with noisy and incomplete annotations [145].

## V. CHALLENGES AND SOLUTIONS

### A. Challenges with Dataset

*1) Shortage of large-scale, annotated, and standardized datasets:* High-quality and large amounts of data are crucial for deep neural network (DNN) models. However, obtaining data, especially sensitive information like medical images, can be difficult through crawling methods. Additionally, manual annotations on medical images can be time-consuming, costly, and require specialized knowledge.

Solutions to address the shortage of datasets include: (a) utilizing unsupervised learning techniques, such as semi-supervised learning, weak supervision, active learning, and transfer learning (as discussed in Section IV4); (b) utilizing open-source datasets, including those mentioned in previous sections, as well as others listed in "10 Open Repositories for the Medical Image Processing Community" by vin bigdata [146]; (c) creating simulated data using the Synthetic Minority Oversampling (SMOTE) technique, which generates new data points based on the closest data points of the minority class; (d) implementing data augmentation [51]; (e) and Deploying Pre-Trained DNN models [147].

*2) Class Imbalance in Datasets:* Medical datasets are often highly imbalanced with regards to class distribution, with some diseases having significantly more negative samples than positive samples, sometimes up to a 98% to 2% ratio. This level of imbalance can result in biased outcomes, and the prediction model may not be accurate for the minority class as it is frequently biased towards the majority class.

Solutions to class imbalance in medical datasets include adjusting the evaluation metric, under-sampling, over-sampling, collecting more data, incorporating model penalties, and using cross-validation.

In addition to class imbalance, medical image datasets can also face issues of sparse annotations and intensity inhomogeneities, which can affect model performance. There are specific solutions for each of these cases presented in the papers [148] and [149].

### B. Challenges with DNN

*1) Training time:* Training a deep neural network (DNN) requires significant time and effort to extract the necessary features and enhance performance.

Solutions to improve the efficiency of DNN training include: (a) implementing batch normalization; (b) using pooling layers to reduce image size and the number of parameters, and employing the early stopping technique to minimize unnecessary training time and avoid overfitting.

*2) Overfitting:* Deep learning algorithms are susceptible to overfitting, where a model becomes too tailored to the training data and memorizes both meaningful patterns and random noise and fluctuations. This can result in poor generalization and low performance on unseen test data.

To prevent overfitting, various techniques can be utilized, as outlined in [150]. These include training with more data, implementing data augmentation, adding noise to input data, performing feature selection, utilizing cross-validation, simplifying data, applying regularization, using ensembling methods, employing early stopping, and incorporating dropout layers.

*3) Gradient vanishing:* Gradients tend to decrease in magnitude as back propagation be activated in a deep neural

network (DNN). This means that the updates performed by Gradient Descent have limited impact on the weights of these layers, making convergence difficult and potentially leading to poor performance of the DNN. This phenomenon is referred to as the Vanishing Gradients problem.

To address the Vanishing Gradients problem, there are two main approaches: (a) Preprocessing, appropriate activation function selection, and proper weight initialization. (b) Using residual blocks and skip connections along the encoder-decoder path, as described in [151].

*4) Computational complexity:* Deep learning algorithms often involve complex calculations, and the training process is typically performed on high-performance hardware such as GPUs or supercomputers.

To address this, it is important to construct compact and portable models with a reduced number of parameters while still maintaining high performance for the intended task [152]. This can be achieved through various techniques such as regularization, model pruning, and efficient architectures

*5) The "black box" nature of AI models:* The "black box" nature of artificial intelligence raises important ethical, transparent, and explainability concerns in clinical medical applications. Despite numerous efforts in the field of XAI, this problem remains a significant challenge for researchers to address.

To address the challenges of explainability in AI, there are several solutions available, beyond those mentioned in Section III, such as CAM, LIME, and SHAP. Other measures include SAU-Net [153], saliency map, and others.

Building on the progress made by the XAI community, Seibold et al. [154] have proposed an indirect segmentation method that leverages Layer Relevancy Propagation (LRP) to extract binary segmentation maps at the pixel level. This approach demonstrates comparable results to UNet while only using image-level labeled data.

*6) Early segmentation problems:* This approach represents a new and promising opportunity for application in clinical settings. It has the potential to drive advancements in both medicine and AI, and ultimately contribute to better protection of human health. This is a significant breakthrough that holds great promise for the future.

Solutions:

In the early stages of tumor development, it can be difficult to detect small tumors due to their size. To address this challenge, it is important to build highly sensitive models that are capable of accurately segmenting small regions of interest, such as the TransUNet+ model described in [50].

To predict the progression of a disease in patients, one approach is to conduct a follow-up study that combines previous and current results to forecast future segmentation outcomes. In practical experience, malignant tumors often grow abnormally and out of control, making early and accurate detection and segmentation critical for reducing the risk of patient mortality. To achieve this goal, it is crucial to develop early and precise auto-segmentation technology. The use of past observations to predict future events is a key method for making informed decisions, as described in [155].

## VI. Conclusions and Future Work

Over the past decade, AI-assisted automated medical image segmentation has gained significant attention from the computer vision community. This study provides a comprehensive overview of the state-of-the-art solutions in terms of network architecture, data, loss functions, and performance metrics. With a focus on network architecture, this research categorizes the solutions to three levels of intelligent vision systems. The study also considers the "intelligence" level of the DIKIW hierarchy.

The paper also highlights two issues that are currently gaining significant attention, which are Explainable AI (XAI) and Medical Image Segmentation (MIS)-based early prediction. Both of these topics are considered hot research areas due to their practical and humanistic values. By addressing XAI and early prediction, the level of the DIKIW hierarchy can be raised from "intelligence" to "wisdom". Despite some progress in these areas, there are still numerous challenges that remain for researchers to tackle.

The next step in the research direction of this paper is to develop a trusted XAI-based early diagnosis support system that can be used in hospitals by both patients and doctors.

The aim of the paper is to advance the use of AI in disease diagnosis to a level of wisdom and, as a result, improve the well-being of society.

### Acknowledgment

The project is supported by the Department of Computer Vision and Cognitive Cybernetics, Faculty of Information Technology, VNUHCM-University of Science.